\documentstyle[aps,prb,epsf,twocolumn]{revtex}

\begin{document}
\draft
\twocolumn[\hsize\textwidth\columnwidth\hsize\csname @twocolumnfalse\endcsname
\begin{title}
{The RKKY interactions and the Mott Transition}
\end{title} 
\author{
G. Moeller$^1$ , V. Dobrosavljevi\'{c}$^2$, and A. E. Ruckenstein$^1$}
\address{$^1$Serin Physics Laboratory, Rutgers University,  
Piscataway, NJ 08855-0849\\
$^2$National High Magnetic Field Laboratory, Florida State 
University\\1800 E. Paul Dirac Dr., Tallahassee, Florida 32306}
\maketitle
\begin{abstract}
A two-site cluster generalization of the Hubbard model in large
dimensions is  
examined in order to study the role of short-range spin correlations near 
the metal-insulator transition (MIT). The model
is mapped to a two-impurity Kondo-Anderson model in a self-consistently 
determined bath, making it possible to directly address the competition 
between the Kondo effect and RKKY interactions in a lattice context. 
Our results indicate that the RKKY interactions lead to qualitative
modifications of the MIT scenario even in the absence of long range 
antiferromagnetic ordering. 
\end{abstract}
]
\newpage
\section{Introduction}
The competition between the Kondo effect and the RKKY interactions
is a recurring theme in many of the most interesting phenomena 
associated with the physics of strong electronic correlations. 
When the RKKY interactions predominate, the result is long-range 
magnetic ordering, as found in many heavy-fermion materials
\cite{heavyferm}. 
In situations where magnetic ordering is absent, the manifestations
are more subtle, but often equally fundamental. In particular, it has
been suggested \cite{colgan} that this competition lies at the core of
the proposed ``two-fluid'' behavior, and ``micromagnetism'' found in
some non-magnetic heavy fermion systems.  Another interesting class of
systems where both the Kondo effect and the RKKY correlations are
believed to be crucial are exemplified by doped semiconductors
\cite{miko} near the metal-insulator transition.  In these systems,
non-Fermi liquid \cite{bf,dkk} metallic behavior is observed,
suggesting the coexistence of local moments and conduction electrons
that seem decoupled from each other -- another manifestation of the
``two-fluid'' behavior
\cite{twoflu}. 
There are many further examples where these effects are of key 
importance. Unfortunately, there are very few theoretical approaches
that are able to treat both the Kondo physics and the RKKY correlations 
on the same footing and provide a convincing picture of these interesting 
phenomena. 

Theoretically, much of our current understanding of strongly
correlated metallic phases relies on a variety of mean-field
descriptions, most of which essentially emphasize the Kondo aspect of
the problem. Several approaches have been proposed, but the most
elaborate one, combining many of the pre-existing ideas in the field,
is based on taking the limit of large spatial dimensionality
\cite{MetzVoll}.  This method represents a generalization \cite{GKS}
of simple, but physically transparent mean-field ideas of Bragg and
Williams \cite{bragg}, as applied to interacting electronic
systems. In this picture, the electron residing on a given site is
viewed \cite{GeoKot} as a Kondo spin which is coupled by an exchange
interaction to a bath consisting of the remaining electrons.
Formally, the problem is mapped \cite{GeoKot} onto an Anderson
impurity model supplemented by an additional self-consistency
condition. The Kondo resonance of the impurity model maps to the heavy
quasiparticle band, and the Mott transition
\cite{Jarrell1,Marcelo1,Marcelo3,GeoKra} is driven by the
vanishing of the Kondo temperature $T_{Kondo}\sim m/m^*$.  Besides
providing an appealing, physically transparent picture of the
correlated state and the metal-insulator transition, the $d=\infty$
method provides a quantitatively accurate computational approach valid
in the entire temperature range. As it treats at the same level both
the coherent, quasiparticle excitations, and the incoherent collective
inelastic processes, the method even allows for a description of fully
incoherent, non-Fermi liquid metallic states
\cite{Si}. 
 
In spite of the successes of the $d=\infty$ mean-field approach, it
remains unable to address several important physical questions. Since
it based on a mapping on a single-site (impurity) model, it cannot
properly account for the competition between the Kondo effect and the
spin-spin correlations between neighboring sites -- the effect that we
have argued is crucial in a number of physical situations. The
locality inherent in this formulation leads to another feature that is
likely to be an artifact of mean-field theory: the ``pinning'' of the
density of states at the Fermi level
\cite{Mueller-Hartmann}.
More precisely,
this effect can be directly traced to the momentum independence
\cite{MetzVoll} of the local 
self-energy, reflecting the lack of spatial correlations. In the
context of strongly correlated, but weakly disordered systems
\cite{dk}, the pinning condition was shown to result in a
discontinuous jump of the DC conductivity at $T=0$ -- the minimum
metallic conductivity. If the pinning is relaxed, it is conceivable
that a continuous behavior of the conductivity would follow, thus
qualitatively modifying our picture of transport near the
metal-insulator transition \cite{miko}.

In order to address the limitations of the existing $d=\infty$ theory,
a most straightforward approach would be to investigate systematic
$1/d$ corrections resulting from finite dimensionality. Several
different methods for performing such expansions have been proposed
\cite{dk,1overd}, but each of these approaches result in formidable
technical difficulties, making it difficult to address the finite
dimensional effects in a simple and elegant fashion. In this paper, we
take an alternative route: we propose to extend the existing theories
{\em in} $d=\infty$ in a way that mimics the most important physical
effects of finite dimensionality. Given the fact that the general
large-dimensions philosophy is based on the mapping of a lattice
models onto appropriate impurity models, the appropriate impurity
model displaying the relevant physics is the two impurity Kondo
(Anderson) model \cite{twoimp}, which is often used as a simplest
model for the study of the RKKY-Kondo competition.  Using standard
methods \cite{GKS}, we can obtain a {\em lattice version} of this
model by self-consistently embedding it in an appropriate medium.  The
resulting model is the ``minimum model'' that allows us to go beyond
the limitations imposed by the conventional $d=\infty$ approach,
without performing uncontrolled or unjustified approximations.

In the rest of this paper, we define and examine this model, and 
indicate how the new features inherent to the RKKY-Kondo competition
modify the standard $d=\infty$ results for the Hubbard model
\cite{Jarrell1,Marcelo1,Marcelo3,GeoKra}.
Specifically, we investigate the modifications of the Mott transition
in a single-band Hubbard model at half filling.  We conclude that the
RKKY interactions represent a {\em relevant} perturbation, relaxing
the pinning condition and qualitatively modifying the nature of the
metal-insulator transition.

\section{The Model}

\begin{figure}
\epsfxsize=3.2in \epsfbox{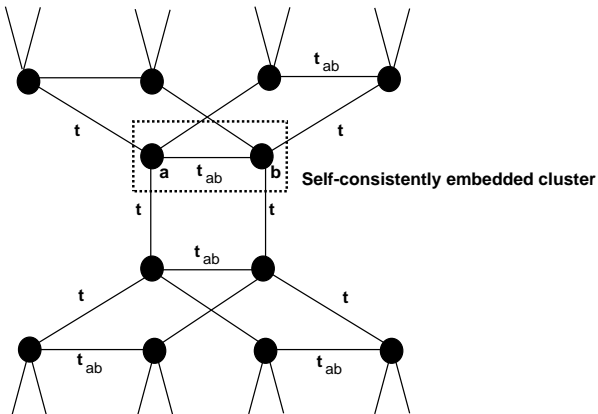}
\caption{
Lattice structure of the doubled Bethe lattice and effective
two impurity cluster.\hfill}
\label{TwoImpLatt}
\end{figure}

We begin our discussion by defining the model that we consider, and
derive the corresponding self-consistency conditions by performing the
$d\rightarrow\infty$ limit.  While the limit of infinite dimensions
does not impose any restrictions on the lattice structure studied
\cite{GKS}, the equations become particularly simple and easy to
derive in the case of a Bethe lattice
\cite{Marcelo1,Marcelo3,dk}.  The {\em qualitative} features
of the model will be identical as on other lattices, and the resulting
spectral functions are closer to the three dimensional situation than
for example on the $d=\infty$ hypercubic lattice.  The ``minimum
model'' that we propose is then obtained by {\em doubling} the Bethe
lattice (with hopping $t$), and allowing the electrons to hop {\em
between} the Bethe lattices with hopping $t_{ab}$. The geometry of the
resulting lattice is shown (for coordination number $z=3$) in
Fig. (\ref{TwoImpLatt}).

Denoting the creation operators corresponding to the two Bethe
lattices with $a^{\dagger}_\sigma$ and $b^{\dagger}_\sigma$, the
Hamiltonian can be written as  
\begin{eqnarray}
H&=&- t \sum_{<i,j>,\sigma} (a_{i\sigma}^\dagger a_{j\sigma}
  +                        b_{i\sigma}^\dagger b_{j\sigma})
  + t_{ab} \sum_{i\sigma} (a_{i\sigma}^\dagger b_{i\sigma} + h.c.),
\nonumber \\
 &+& U  \sum_{i\sigma} (n_{ai\sigma} n_{ai-\sigma}
                        +n_{bi\sigma} n_{bi-\sigma}) -\mu
\sum_{i\sigma} (n_{ai\sigma} +n_{bi\sigma})
\end{eqnarray}
where $U$ is the Coulomb potential and $t$ is the nearest-neighbor hopping
amplitude; $t_{ab}$ is the hopping amplitude between the two lattices.

It should be stressed at this point that this model clearly breaks
translational invariance by singling out {\em pairs} of sites 
connected by hopping elements $t_{ab}$. While this feature appears
somewhat artificial in a uniform system in which all neighbors are
equivalent, it leads to a {\em controlled} and non-trivial modification
of the  $d=\infty$ limit \cite{gk2}.
In contrast to the standard single-band
Hubbard model in infinite dimensions, in which
electrons solely undergo {\em temporal} fluctuations, our model also
allows for  {\em spatial} fluctuations. A systematic expansion in
$1/d$ \cite{dk,gk2} includes exactly these processes and the
model can therefore be interpreted as including some of the effects of
finite dimensionality. 

It is clear that the model by construction enables us to study 
nearest-neighbor spin-correlations.
In physical terms, for $t_{ab}$ large, the model favors the formation 
of singlet pairs (dimers) from the ``a-b'' sites. Interestingly, 
this symmetry breaking is not unreasonable in {\em disordered
systems}, where each site  $a$ has another ``preferred'' neighboring
site $b$, with which dimerization will be favored. This notion is at
the heart of the ``random singlet'' ordering of Bhatt and Lee \cite{bl},
describing the singular thermodynamics of doped semiconductors. 
Notice, however, that a variety of additional interpretations is
possible. In particular, the model may alternatively viewed as a two
band model or  as two coupled layers \cite{monien}.

As usual, the problem simplifies considerably in the large
coordination (large dimension) limit, where a mapping to an
appropriate impurity model is obtained. 
Using standard methods \cite{GKS}, we proceed by 
rescaling the hopping amplitude $t$ as $t \rightarrow
\frac{t}{\sqrt{m}}$ ($m=z-1$ is the ``branching ratio'' of the
Bethe lattice), and taking the limit $m\rightarrow \infty$.
The result is an effective {\em two-impurity Anderson model}
embedded in a self-consistently determined bath.
We introduce spinors ${\bf c}^*_\sigma(\tau)
=(a^*_{\sigma}(\tau),b^*_{\sigma}(\tau))$ and the matrix
Green function
\begin{equation}
{\bf G}_\sigma (\tau-\tau')=
\left( \begin{array}{cc} -\langle{\cal T} a_\sigma {\tau} a_\sigma 
{\tau'}^{\dagger}\rangle & 
-\langle{\cal T} a_\sigma ({\tau})
b_\sigma^{\dagger}(\tau')\rangle \\ 
-\langle{\cal T} b_\sigma({\tau}) a_\sigma^{\dagger}(\tau')\rangle &
-\langle{\cal T} b_\sigma({\tau}) b_\sigma^{\dagger}(\tau')\rangle 
\end{array} \right),
\label{MatG2}
\end{equation}
with
\begin{equation}
{\bf G}(i\omega_n)=-\int_0^{\beta} e^{i \omega_n \tau} \langle T_\tau
{\bf c}(
\tau) {\bf c}^{\dagger}(0) \rangle_{S_{eff}}.
\label{DefG}
\end{equation}
Notice that due to spin conservation
${\bf G}_{\sigma} = \delta_{\sigma \sigma '}  {\bf G}_{\sigma\sigma '}$.

The effective action can then be written in matrix form as
\begin{eqnarray} 
S_{eff} [{{\bf c_{\sigma}}},{{\bf c_{\sigma}^*}}]
&=&-\sum_{\sigma}\sum_{i \omega_n} 
{{\bf c_{\sigma n}^*}} {{\bf G_{0}^{-1}}}(i \omega_n) {{\bf c_{\sigma n}}}
\nonumber \\& & + U \int_0^{\beta} d\tau 
(n_{a \uparrow}n_{a \downarrow}+n_{b \uparrow}n_{b \downarrow})
\label{EffAc2}
\end{eqnarray}
where the self-consistency condition reads
\begin{equation}
{\bf G_{0}^{-1}}(i \omega_n)= \left(
\begin{array}{cc} i \omega_n+\mu & -t_{ab}\\
                  -t_{ab}    & i \omega_n+\mu \end{array} \right) - t^2
{\bf G(i \omega_n)}.
\label{SC2}
\end{equation}

While solving this model for general values of the parameters
represents a highly nontrivial task, we immediately recognize some
well known limiting cases. In the limit $t_{ab}=0$, the two Bethe
lattices decouple, and the model reduces to the well known single-band
Hubbard model in $d=\infty$
\cite{Jarrell1,Marcelo1,Marcelo3,GeoKra}.  At half filling,
this model undergoes a Mott transition at $U=U_{c_2}$, which is
preceded by a formation of a pseudo-gap and the coexistence
\cite{Marcelo1,MarMoe} of a metallic and an insulating solution in the
region $U=U_{c_1} < U < U=U_{c_2}$. However, the metallic solution is
lower in energy \cite{MarMoe} at $T=0$ throughout the coexistence
region, so that $U=U_{c_2}$ represents a true zero-temperature
critical point where the two solutions merge.

The other easily analyzable case is the noninteracting limit $U=0$. 
Here, a band-crossing transition takes place, where the density
of states (DOS) at the Fermi level vanishes continuously and a gap
opens at $t_{ab}=t$. The origin of this transition is easy to
understand: in the atomic limit $t_{ab} >> t$ the DOS reduces to the
two (bonding and antibonding) levels at $E_{\pm} =\pm t_{ab}$. When
the hopping $t$ increases, these atomic levels broaden into bands of
width $\approx 2t$, so that the gap closes when the two bands
overlap, at $t_{ab} =t$. 

\section{Results}

While the limit of infinite dimensions simplifies the original problem
considerably, solving the corresponding impurity model
is still a formidable task. In the framework of the single-band
Hubbard model, a number of numerically exact techniques 
\cite{Jarrell1,Marcelo1,Marcelo3,MarMoe,CafKra}, as well as simpler
approximate methods \cite{GeoKot,Marcelo1,Marcelo3} have been used.
A particularly simple approach first proposed by Georges and Kotliar
\cite{GeoKot} 
is based on solving the Anderson impurity model using second order
perturbation theory, following Yamada and Yosida
\cite{Yamada1,Yamada2,Yamada3}. 
Due to the additional self-consistency  this approach,
often called the ``iterated perturbation theory'' (IPT)
\cite{GeoKra}, still has non-perturbative character.
It is exact in both limits of $U=0$ and $U=\infty$ and displays a Mott
metal-insulator transition.  Detailed investigations based on other
numerical approaches demonstrated \cite{CafKra,MarMoe} the qualitative
validity of most IPT predictions for the single-band Hubbard model in
$d=\infty$.  As compared to numerically exact solutions, IPT requires
considerably less computational effort, and thus represents a valuable
guide to the physics of $d=\infty$ electrons.

\begin{figure}
\epsfxsize=3.2in \epsfbox{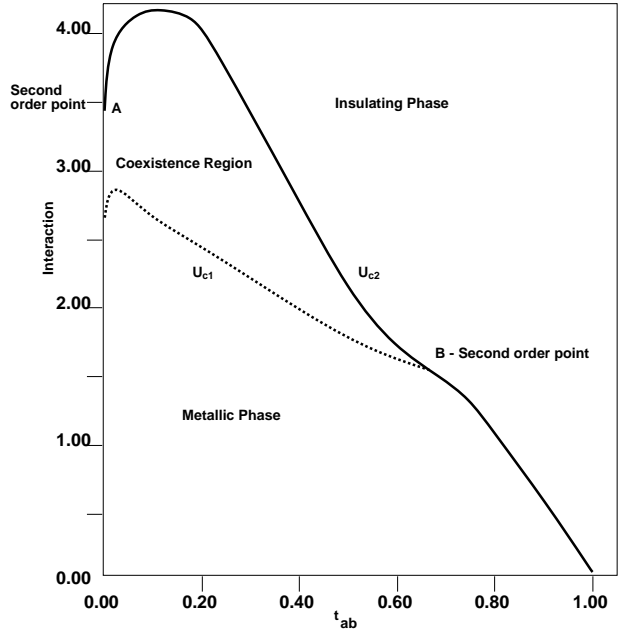}
\caption{
Phase diagram in the $U-t_{ab}$ plane. Both the metallic and the 
insulating solution are locally stable in the coexistence 
region even at $T=0$, in contrast to the $t_{ab} =0$ situation.
The curves denoted by $U_{c1}$ (dashed line)and $U_{c2}$ 
represent the boundaries (spinodals) of the insulating and 
the metallic solution, respectively. The two solutions {\em 
merge} at the two second order (critical) points denoted by $A$
and $B$. A direct, continuous transition from metal to insulator
is found along the critical line (to the right of point B), which 
is qualitatively similar as for $U=0$ (see Figs. 3 and 4). 
\label{PhaseDiag}\hfill}
\end{figure}

In the problem that we consider in this paper, one has to solve a two
impurity Anderson model -- a task which is considerably more difficult
than the simpler one impurity model. Furthermore, numerical Monte
Carlo approaches \cite{hirsch} to the two impurity Anderson model have
proven to be largely unsuccessful at the available computational
level.  Alternative, exact diagonalization approaches also appear
hardly feasible in the case of the two impurity problem, although
recent developments \cite{project} hold considerable promise for the
near future.  Taking these facts into consideration, we propose to
begin the investigation of the problem considered using the IPT
approach as a useful first attempt to gain insight into the RKKY-Kondo
competition.  We note however, that in contrast to the earlier
application of the IPT approach, in the present case this
approximation is not exact in the $U=\infty$ limit, even at half
filling. Still, we do not expect that these limitations will
qualitatively modify our conclusions, especially in view of the
absence of a small energy scale at the first order metal-insulator
transition that we find. The possible instances where the limitations
of the IPT approach could be relevant will be further discussed in
Section IV, where we also present a quantitative estimate for the
range of its validity.

In the following, we will concentrate on the behavior at half-filling,
where the Mott transition takes place at $t_{ab} =0$, and investigate
the modifications induced by turning on $t_{ab}\neq 0$.  In order to
apply IPT to the present model, we have to compute the second order
perturbation theory corrections around the non-magnetic Hartree-Fock
solution.  The second-order diagonal/off-diagonal self-energies in
this case consist of only one diagram respectively, and are given as

\begin{equation}
\Sigma_{xy}(\tau)=-U^2 G_{xy}^0(\tau) G_{xy}^0(-\tau)
G_{xy}^0(\tau),
\end{equation}
where  $x,y=a,b$. 
Since the resulting  equations have to be solved self-consistently,
the solution is obtained by numerical iteration until convergence
is found. 

\begin{figure}
\epsfxsize=3.2in \epsfbox{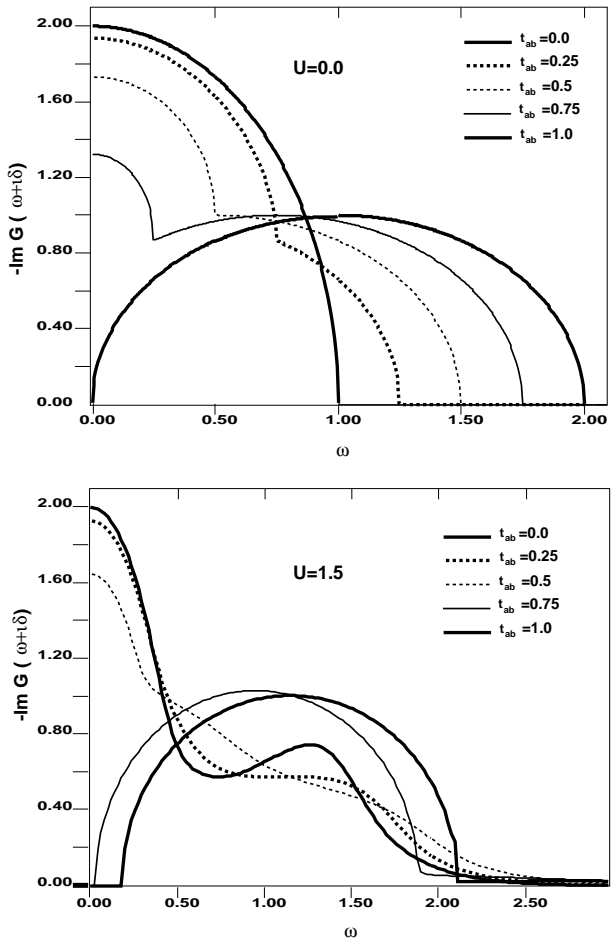}
\caption{
Density of states for interactions a) $U=0$ and b) $U=1.5$ for
$t_{ab}=0.0,0.25,0.5,0.75,1.0$.
\label{DosLow}\hfill}
\end{figure}

We have determined the $T=0$ phase diagram of the investigated model
at half filling using the IPT approximation, and the results are
presented in Fig. (\ref{PhaseDiag}). At small values of $U$, as
$t_{ab}$ is increased, the behavior is qualitatively the same as at
$U=0$. A {\em continuous} transition takes place, at a critical value
of the hopping $t_{ab}^c (U)$ that is found to decrease as $U$
increases.  This behavior reflects the fact that the gradual band
broadening due to the Hubbard-Mott splitting tends to close the gap.
As an illustration we display the evolution of the DOS as the
transition is approached in Fig. (\ref{DosLow}) at $U=0$
(Fig.\ref{DosLow}(a)) and $U=1.5$ (Fig. \ref{DosLow}(b)). [Throughout
the paper, all energies are measured in units of the half-bandwidth
$D=2t$]

\begin{figure}
\epsfxsize=3.2in \epsfbox{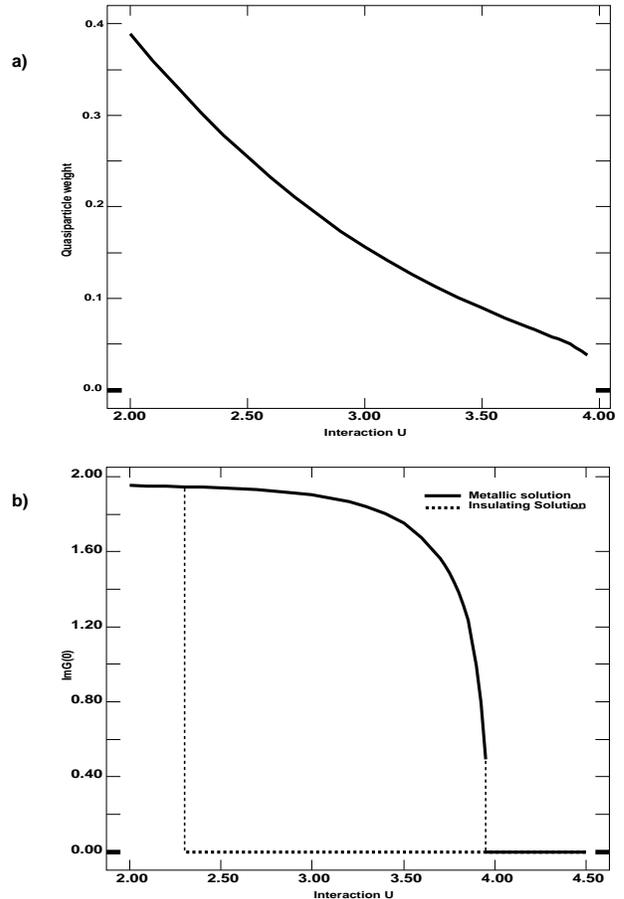}
\caption{
a) Quasiparticle residue $z=1/(1-\frac{\partial
\Sigma}{\partial \omega})$ in the metallic phase as a function of the
interaction $U$ for $t_{ab}=0.2$. b) Density of States ($-Im G(0^+)$)
of metallic (solid line) and insulating (dotted line) solutions at
$t_{ab}=0.2$ as a function of the interaction $U$.
\label{KondoT}\hfill}
\end{figure}

\begin{figure}
\epsfxsize=3.2in \epsfbox{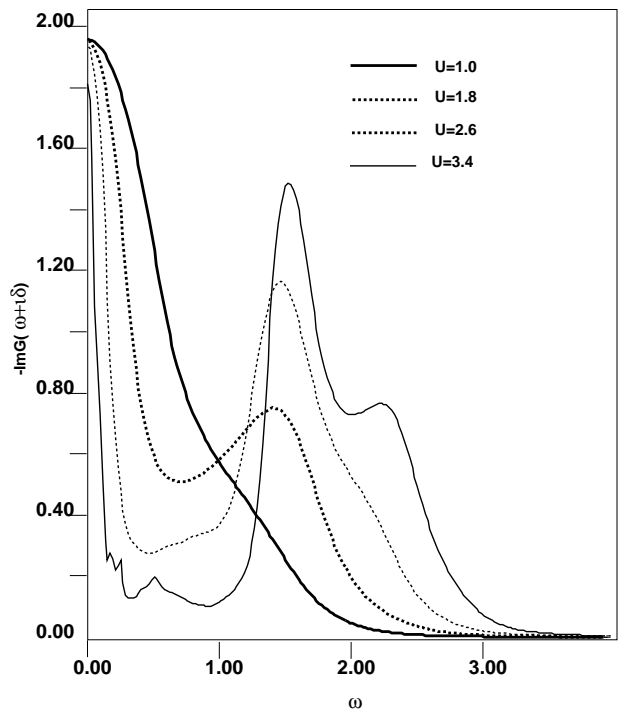}
\caption{
Density of States for $t_{ab}=0.2$ and interactions
$U=1.0,1.8,2.6,3.4$.
\label{DosTrans}\hfill}
\end{figure}

For larger values of $U$, a coexistence region of the
metallic and insulating solutions is found, similarly as for $t_{ab}
=0$. The metallic solution is found for $U<U_{c_2} (t_{ab})$; 
we note the non-monotonic dependence of $U_{c_2} (t_{ab} )$, which is
first found to increase, and then to decrease as a function of
$t_{ab}$. Thus, as compared with $t_{ab}=0$, the addition of the RKKY 
correlation is found to extend the metallic region. More importantly,
we find that the metallic solution disappears {\em discontinuously} at 
$U=U_{c_2} (t_{ab} )$, in contrast to the $t_{ab}=0$ behavior
\cite{Marcelo1,Marcelo3}.
The solution along this boundary is {\em not} characterized by a low energy
scale, as seen by plotting the Kondo temperature (quasiparticle weight
$z=1/(1-\frac{\partial \Sigma}{\partial \omega})$) on
one site as a function of $U$, for $t_{ab}=0.2$ (Fig.
\ref{KondoT}(a)). The density of states is also discontinuous at this
boundary (Fig.\ref{KondoT}(b)).  We note the pronounced dependence of
the density of states at the Fermi energy on the interaction $U$ -- a
clear violation of the ``pinning condition'' that is observed at
$t_{ab} =0$.  This behavior is seen even more clearly by plotting the
evolution of the metallic DOS as the boundary is approached in Fig.
(\ref{DosTrans}).  Clearly, in contrast to the situation at
$t_{ab}=0$, the metallic and insulating solutions {\em do not merge}
at $U=U_{c_2} (t_{ab} )$, so this boundary cannot be identified with a
critical line.

Similar behavior is obtained by examining the stability of the
insulating solution, which is found to {\em discontinuously }
disappear at $U=U_{c_1} (t_{ab})$. As we can see from the phase
diagram, Fig. (\ref{PhaseDiag}), the boundaries $U_{c_1} (t_{ab} )$
and $U_{c_2} (t_{ab})$ are found to join at the critical point ``B''
which is also the end of the band crossing transition critical line.

We thus find that in the entire coexistence region, the metallic and
the insulating solutions merge only at two points: A ($t_{ab} =0$,
$U=U_{c_2}$) and B. In the rest of the phase diagram the two solutions
are {\em disjoint} from each other, and the transition has a {\em
first-order} character. This conclusion can be established even more
rigorously by examining the {\em local stability } of each solution
throughout the coexistence region. For this purpose, we have developed
an approach that allows us to determine the stability, as described in
detail in the appendix. Using this method, we have established that
both the metallic and the insulating solution {\em are locally
stable}, supporting the first-order scenario. (Note that when the
procedure is applied in the $t_{ab}=0$ limit, we find that in the
coexistence region, the insulating solution is locally {\em unstable}
with respect to the metallic solution, in agreement with well
established results
\cite{Marcelo3,MarMoe}). 

\begin{figure}
\epsfxsize=3.2in \epsfbox{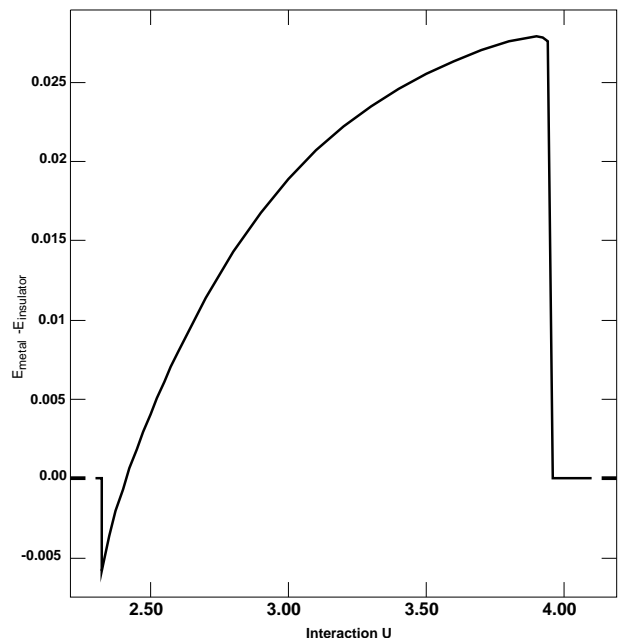}
\caption{
Difference between the energies of metallic and insulating
solutions for $t_{ab}=0.2$ as a function of the interaction $U$.
\label{EnergiesTwo}\hfill}
\end{figure}

To obtain the location of the transition line, we have calculated the
energies of the solutions, and determined the line where they
coincide, as shown in Fig. (\ref{PhaseDiag}).  Since the solutions
merge at points A and B (see Fig. (\ref{PhaseDiag})), the energies of
the solutions have to coincide there, and the (first order) transition
line connects those two points.  As an illustration, the energies of
the two solutions are plotted for $t_{ab}=0.2$ as a function of $U$ in
Fig. (\ref{EnergiesTwo}).  As we can see, in contrast to the $t_{ab}
=0$ findings, the {\em insulating } solution is lower in energy for
larger values of $U$, consistent with the first-order scenario. This
result is perhaps not surprising, as the RKKY interactions are
generally expected to stabilize the insulating solution.

In line with this first order scenario, the boundary lines $U_{c_1}
(t_{ab} )$ and $U_{c_2} (t_{ab} )$ should be recognized as {\em
spinodal} lines. An interesting question is why the two solutions
merge at $t_{ab}=0$, i. e. why is there a critical point there instead
of a first-order transition. The existence of {\em bifurcations},
i. e. critical points is usually associated with spontaneous breakdown
of some symmetry (e.g. up-down symmetry in the Ising model). In the
case of the single-band Hubbard model in $d=\infty$ the relevant
symmetry remains yet to be discovered.

\begin{figure}
\epsfxsize=3.2in \epsfbox{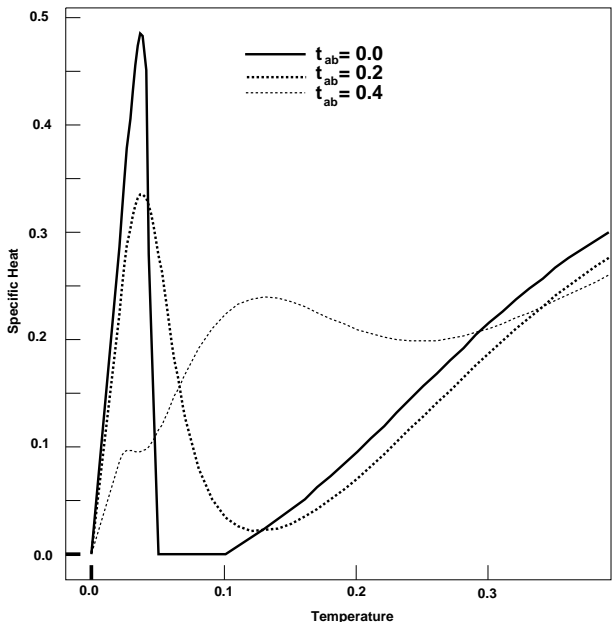}
\caption{
Specific heat as a function of temperature for the metallic
solution, at U=2.5 for $t_{ab}=0.0,0.2,0.4$.
\label{SpeheMet}\hfill}
\end{figure}

The effects of the RKKY interactions are not limited to the
modifications of the MIT scenario. They can also modify the
thermodynamic behavior by affecting the dynamics of the collective
spin fluctuations governing the finite temperature response. In order
to investigate this aspect of the problem, we have computed the
specific heat in the metallic and the insulating phases of our
model. To illustrate the typical metallic behavior, we present results
for the specific heat at $U=2.5$, for three different values of
$t_{ab} =0,\; 0.2,\;0.4$ in Fig. (\ref{SpeheMet}).

At $t_{ab}=0$ we recognize the characteristic linear specific heat at
$T<<T_{Kondo}\sim 0.05 $, corresponding to Fermi liquid behavior, a
Schottky-like peak at $T\sim T_{Kondo}$ reflecting the binding energy
of the Kondo singlet, and insulating-like behavior at $T\sim U/2$ due
to charge fluctuations (Hubbard bands).

For $t_{ab}\neq 0$, i.e. as the RKKY interactions are introduced, a
{\em new} feature appears in the intermediate energy range: the
specific heat is enhanced at $T\sim J_{ab}$, reflecting the emergence
of additional spin fluctuations with a characteristic energy
corresponding to the RKKY exchange interaction $J_{ab}\approx
4t_{ab}^2 /U$. For $t_{ab}=0.4$ and $U=2$ we estimate $J_{ab}\approx
0.08$, giving a (``Schottky'') peak in the specific heat at $T_J\sim
J/2\approx 0.16$, exactly where the observed enhancement occurs.

\begin{figure}
\epsfxsize=3.2in \epsfbox{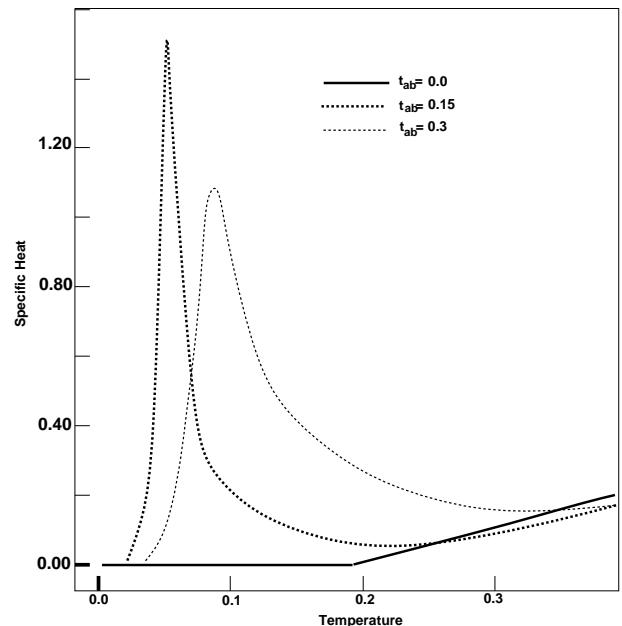}
\caption{
Specific heat as a function of temperature for insulating solution,
 at U=3.5 for $t_{ab}=0.0,0.15,0.3$.
\label{SpeheIns}\hfill}
\end{figure}

This interpretation of the observed specific heat enhancement finds
additional support by examining the corresponding behavior in the
insulating regime. Typical results are presented in Fig.
(\ref{SpeheIns}), where the specific heat is plotted for $U=3.5$, and
$t_{ab} =0,\; 0.15$ and $0.3$.

In absence of RKKY interactions ($t_{ab}=0$) the specific heat is
vanishingly small at low temperatures, reflecting the existence of the
Mott-Hubbard gap. As in the metallic phase, the addition of RKKY
interactions ($t_{ab}\neq 0$) induces specific heat enhancements in a
comparable temperature range, at $T_J\approx 2t_{ab}^2/U$.  We also
note that the corresponding enhancement in the metallic state (Fig.
\ref{SpeheMet}) is much more spread-out in temperature, presumably
reflecting the scattering of these fluctuations by the coupling to the
particle-hole excitations. 

\section{Beyond Perturbation Theory}
The solution of our model presented in the preceding section was based
on an approximate scheme for the impurity problem -- the perturbation
theory approach of Yosida and Yamada \cite{Yamada1} (YY). While this
techniques was utilized with impressive success in previous $d=\infty$
studies \cite{GeoKot,GeoKra,Marcelo1}, it is important to emphasize
the limitations of this approach, and identify instances where most
important problems can be expected.

When applied to single-impurity Anderson models, the approach of YY is
generally expected to be at least qualitatively correct in the entire
temperature range. In this case, the ground state is a (local) Fermi
liquid, so that perturbation theory converges \cite{Horvatic} and
finite order corrections are sufficient. The situation is more
complicated in two-impurity models such as the two-impurity Kondo
(Anderson) model.  Here, a critical point \cite{twoimp} is found at
half-filling, separating the RKKY and Kondo regimes. The emergence of
this critical point has a simple physical origin. It reflects the fact
that two ground states with different symmetry are possible,
corresponding to the Kondo spins being compensated either by
conduction electrons (Kondo phase) or by each other (RKKY phase). This
critical point, which reflects a degeneracy due to level crossing
\cite{rasul} signals a breakdown of a Fermi liquid description. As a
result, we do not expect perturbative approaches to be accurate in the
critical region. Indeed, if the two-impurity problem is treated in
perturbation theory, the critical point is washed out into a smooth
crossover.

Without providing a more elaborate treatment of the two-impurity
problem, we can at least make estimates of the regions in parameter
space where IPT could prove insufficient. Based on the information
available from studies of the two impurity Kondo model \cite{twoimp}
in a fixed bath, we expect that a critical point emerges when the RKKY
interaction $J_{ab}$ is comparable to the ``bare'' Kondo temperature
$T_{Kondo}^o =T_{Kondo} (t_{ab}=0 )$. Since near $U_{c_2}$ the Kondo
temperature vanishes
\begin{equation}
T_{Kondo}^o \sim \frac{m_o}{m^*}\sim (U_{c_2} -U),
\label{TKondo}
\end{equation}
but the exchange interaction remains finite
\begin{equation}
J_{ab}\sim t_{ab}^2 /U
\label{JExch}
\end{equation}
one can expect that increasing $U$ at finite $t_{ab}$ drives the
system from a Kondo to an RKKY metallic phase. The critical line where
this could take place can be estimated by equating $T_{Kondo}^o$ and
$J_{ab}$, and using Eqs. (\ref{TKondo})-(\ref{JExch}), we find
\begin{equation}
U_{RKKY} (t_{ab} )\approx U_{c_2} -4t_{ab}^2/U_{c_2}.
\end{equation}

\begin{figure}
\epsfxsize=3.2in \epsfbox{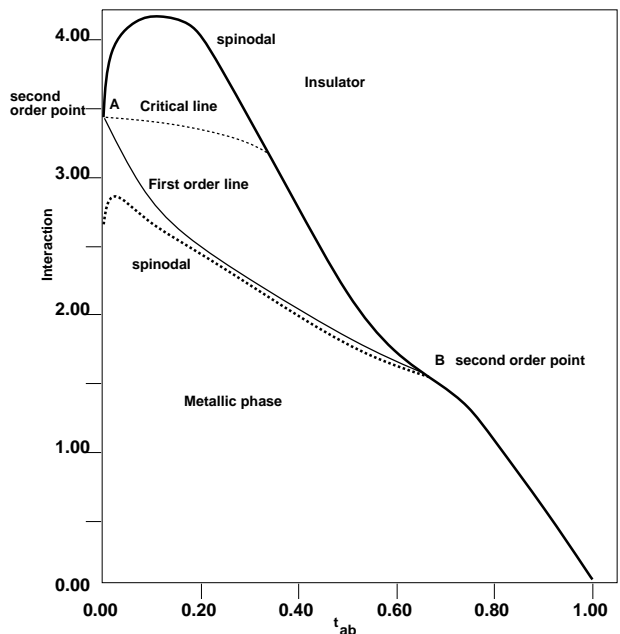}
\caption{
Phase diagram indicating the critical line corresponding to
the Kondo-RKKY phase transition. Also shown is the first order line
found in our model.
\label{EstimBound}\hfill}
\end{figure}

This expression is valid only in the $t_{ab}\rightarrow 0$ limit,
where to leading order we have ignored the modifications of the
(self-consistently adapting) electronic bath. This estimate is plotted
in Fig. (\ref{EstimBound}), where it is compared with the perturbation
theory predictions for the metallic phase boundary, and the location
of the first-order metal-insulator transition.

As we can see, according to IPT, the first order transition {\em
preempts} the approach to the RKKY-Kondo critical line, supporting the
validity of IPT-based predictions.

The ITP prediction that the introduction of RKKY interactions induces
a first-order metal insulator transition finds additional support if
we recall that a similar conclusion was obtained by introducing
additional RKKY interactions in the large-N approaches to correlated
electrons \cite{jterm}.  However, we emphasize that this approach did
not have the two impurity Kondo physics build in, and, in particular,
the possibility that the RKKY-Kondo competition induces a nontrivial
critical point even on the impurity level.

An interesting question that deserves further study is the role of the
two impurity Kondo model critical point in the destruction of the
metallic phase. Of course, this question would be particularly
relevant if additions of small perturbations, perhaps disorder, could
stabilize the metallic phase to larger values of $U$, so that the
relevant critical point becomes physically accessible.

\section{Conclusions}
In this paper, we addressed the role of short-ranged magnetic
correlations in determining the behavior of strongly correlated
electronic systems. To account for these effect, which are not
properly treated by existing approaches, we propose a two-site cluster
generalization of the Hubbard model in infinite dimensions as the
simplest model containing the relevant physics. The model is mapped
onto a two-impurity Kondo-Anderson model in a self-consistently
determined bath, making it possible to directly address the
competition between the Kondo effect and RKKY interactions in a
lattice context.

Using a well known approximation scheme for solving the
self-consistency conditions, we have determined the phase diagram of
our model and discussed the modifications of the metallic behavior.
We find that the addition of RKKY interactions induces a first-order
metal-insulator transition, by energetically favoring the insulating
phase. Additional low-energy spin fluctuations emerge, leading to
enhancements of the specific heat in the intermediate temperature
range, both in the metallic and insulating phases.

\begin{center}
\bf APPENDIX: {STABILITY OF $d=\infty$ SOLUTIONS}
\end{center}

In this appendix, we describe a method that can be used to examine the
local stability of the $d=\infty$ self-consistency equations, Eqs. 
(\ref{MatG2})-(\ref{SC2}). The method is based on the observation that
these equations can be derived using a variational approach, i.e. by
extremizing a certain functional in analogy with standard
Landau-Ginsburg formulations of mean-field theory. 

In the present case, this functional takes a form
\begin{equation}
F[ {\bf G}_{\sigma} (i\omega_n ) ]=
-\frac{1}{2} t^2 \frac{1}{\beta}
\sum_{\sigma}\sum_{\omega_n} tr[{\bf G}^2_{\sigma} (\omega_n )]
+ F_{imp} [ {\bf G}_{\sigma} (i\omega_n ) ],
\label{LandauFunc}
\end{equation}
where $F_{imp} [ {\bf G}_{\sigma} (i\omega_n ) ]$ is the free energy of
the two-impurity model as defined by the action of Eqs. (\ref{EffAc2})
and (\ref{SC2})
\begin{equation}
 F_{imp} [ {\bf G}_{\sigma} (i\omega_n ) ]=
-\frac{1}{\beta}
\ln \int Dc^{*} Dc e^{-S_{eff} [{\bf G}_{\sigma }  (i\omega_n ) ]}.
\label{FreeEn}
\end{equation}
Here, we consider Eqs. (\ref{EffAc2})-(\ref{SC2}) as a {\em
definition} of the effective 
action, so that $ F [ {\bf G}_{\sigma} (i\omega_n ) ]$ is a functional of 
a {\em arbitrary}, yet unspecified function $ {\bf G}_{\sigma} (i\omega_n
)$. 

In analogy with conventional Landau-Ginsburg formulation, we think 
of ${\bf G}_{\sigma }  (i\omega_n )$ as an order parameter (function).
The mean-field equations are then obtained by extremizing the 
above functional with respect to variations in the form of ${\bf
G}_{\sigma }  (i\omega_n )$.  
The extremum condition reads
\begin{equation}
\frac{\delta F[{\bf G}]}{\delta {\bf G}_{\sigma }  (i\omega_n )}=0,
\label{DfDg}
\end{equation}
giving
\begin{equation}
{\bf G}_{\sigma }  (i\omega_n )={\bf G}_{\sigma }^{imp}  (i\omega_n ).
\end{equation}
Here, ${\bf G}_{\sigma }^{imp}  (i\omega_n )$ is the local 
Green function of the impurity model corresponding to a 
{\em fixed } ``bath'' Green function ${\bf G}_{\sigma }  (i\omega_n )$
\begin{equation}
{\bf G}_{\sigma }^{imp}  (i\omega_n )=
\langle c^{*}_{\sigma }  (i\omega_n )\; c_{\sigma }  (i\omega_n
)\rangle_{S_{eff} [G]}.
\label{GEq}
\end{equation}
Note that  ${\bf G}_{\sigma }^{imp}  
(i\omega_n )$ is also a {\em functional }
of $ {\bf G}_{\sigma }  (i\omega_n )$. 

Obviously, Eq. (\ref{GEq}) is identical to Eq. (\ref{DefG}), so that
we recover the $d=\infty$ the self-consistency conditions Eq.
(\ref{MatG2})-(\ref{SC2}).  
 
Before going further, it is worthwhile to comment on the physical
interpretation of the functional of Eqs. (\ref{LandauFunc})-(\ref{FreeEn}). 
Using the expressions for the free energy of $d=\infty$ 
models \cite{GKS}, one can  show that our functional
reduces to the {\em free energy}, when evaluated for the value of 
${\bf G}_{\sigma }  (i\omega_n )$ corresponding to the solutions of 
the self-consistency conditions, Eqs. (\ref{MatG2})-(\ref{SC2}).
We conclude that $F[{\bf G}]$ represents a {\em free energy
functional}, in the usual Landau-Ginsburg sense.

When the self-consistency conditions are solved numerically, one
typically makes an initial guess for $ {\bf G}_{\sigma } (i\omega_n
)$, defining the effective action of the impurity model using
Eq. (\ref{SC2}).  The impurity model is then solved by any method
available, and a {\em new} value of $ {\bf G}_{\sigma } (i\omega_n )$
obtained from Eq.  (\ref{DefG}). In numerical analysis, this iterative
procedure is known as the ``substitution-iteration method'', which is
repeated until convergence is reached. In the following, we prove a
general theorem that such an iterative procedure converges towards a
local (nearest) {\em minimum} of the free energy functional. We note
that the set of all possible functions ${\bf G}_{\sigma } (i\omega_n
)$ form a vector space (more precisely an infinite dimensional Hilbert
space), and for notational simplicity, denote these vectors by ${\bf
x}$.

We define the {\em gradient} vector field ${\bf g}({\bf x} )$ as
\begin{equation}
{\bf g}({\bf x} )=\partial_{{\bf x}} F[{\bf x} ],
\end{equation}
so that Eq. (\ref{DfDg}) can be written as
\begin{equation}
{\bf g}({\bf x} )|_{{\bf x} ={\bf x}_o}=0.
\end{equation}
Here, ${\bf x}_o$ corresponds to the solution of the self-consistency
condition, i.e. is a locally stationary point of $F[{\bf x} ]$. 
If we define further the quantity 
\begin{equation}
{\bf f}= {\bf x} -{\bf g}({\bf x} ), 
\end{equation}
we find that at the stationary point 
\begin{equation}
{\bf x}_o ={\bf f}({\bf x}_o ).
\end{equation}
In this language, the ``substitution-iteration' search for the 
solution can be written as
\begin{equation}
{\bf x} (n+1)={\bf f}({\bf x} (n) ),
\end{equation}
and the solution corresponds to 
\begin{equation}
{\bf x}_o =\lim_{n\rightarrow\infty} {\bf x} (n) .
\end{equation}

Note that the {\em increment} of ${\bf x} (n)$ can be also written as
\begin{equation}
\Delta{\bf x} (n) ={\bf x} (n+1)-{\bf x} (n)\;  =\; -{\bf g}({\bf x} (n)).
\end{equation}
As we can see, the iteration takes the vector ${\bf x}$ in the direction 
opposite to the gradient, i. e. ``down the hill'' , so that the
iteration converges only in the vicinity of a {\em local minimum}. 

On general grounds, we expect the physical solutions near first-order
transitions to be {\em locally stable}. We can check this stability,
by making a small modification in the initial conditions that produce
the respective solutions. More precisely, we should first find the
convergent metallic and insulating solutions ${\bf x}_o^M$ and ${\bf
x}_o^I$.  We can then examine the stability of, for example, the
insulating solution by re-starting the iteration search from a new
initial guess
\begin{equation}
{\bf x}^i (n=0)= (1-c){\bf x}_o^I + c{\bf x}_o^M.
\end{equation}

The solution is locally stable if for $c$ sufficiently small the
iteration procedure converges to ${\bf x}={\bf x}_o^I$. We can
similarly check the stability of the metallic solution by choosing
$c\approx 1$.

In order to apply these ideas, we first test them in the well examined
limit corresponding to $t_{ab}=0$. We find that the metallic solution
is stable throughout the coexistence region, but that the insulating
solution becomes unstable as $T\rightarrow 0$.  These findings are in
complete agreement with the result \cite{MarMoe} that $U_{c2}$ is a
$T=0$ critical point at which (upon reducing $U$) the insulating
solution becomes unstable, and a new metallic solution emerges.  We
can apply these ideas for $t_{ab}\neq 0$, in which case {\em both}
solutions are found to be locally stable, in agreement with a
first-order scenario.

\begin{figure}[t]
\epsfxsize=3.2in \epsfbox{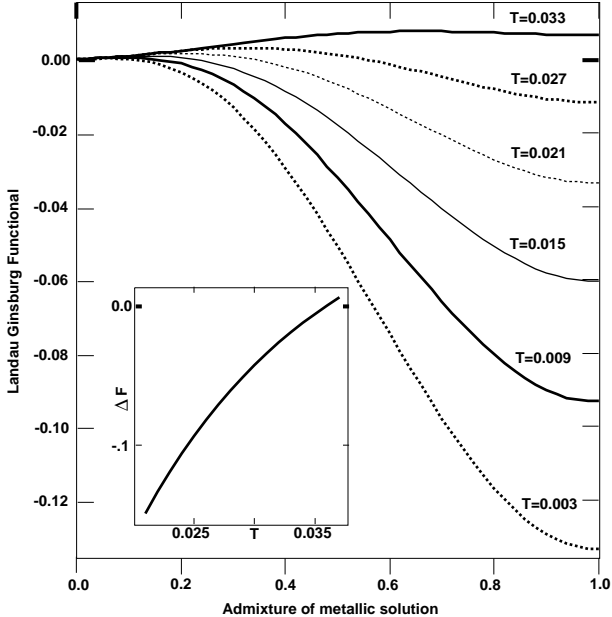}
\caption{
Landau free energy for $t_{ab}=0$, $U=2.6$ and temperatures
$T=0.003-0.033$ in steps of $\Delta T=0.006$. The inset shows the free
energy difference between the two minima as a function of temperature.
\label{Landau}\hfill}
\end{figure}

We conclude this discussion by an explicit construction of the 
``Landau-Ginsburg'' functional, which represents a nice illustration
of the above stability considerations. 
To do this, we note that in the case of the $d=\infty$ equations, the 
gradient vector takes the form
\begin{equation}
{\bf g}={\bf G}^{imp} [{\bf G} ] -{\bf G},
\end{equation}
which can be calculated by any method that solves the Anderson 
impurity model, e.g. the YY approach \cite{Yamada1}. 
Once the gradient is available, 
it is possible to determine the evolution of $F[{\bf x} ]$ along any
particular direction in the ${\bf x}$ space. In particular, we expect the
physical solutions to be local minima,  separated by an unstable
solution (local maximum or a saddle-point). It is thus useful to
consider the direction (vector) connecting the two solutions, which
can be parameterized as
\begin{equation}
{\bf x} (\ell )=(1-\ell ){\bf x}_o^I +\ell{\bf x}_o^M.
\end{equation}
The {\em increment} of $F[{\bf x} (\ell )]$ can be expressed as a line
integral
\begin{equation}
\Delta F(\ell )=F[{\bf x} (\ell )]-F[{\bf x}_o^I ]=
\int_0^1 \; d{\bf \vec \ell}\cdot {\bf g}({\bf x} (\ell )).
\end{equation}
We can numerically compute this line integral by an appropriate
discretization procedure, and typical results in the $t_{ab}=0$ limit
are plotted in Fig. (\ref{Landau}).

Here, we show $\Delta F(\ell )$ for
$U=2.6$ and for several different temperatures. As we can see, at 
$T\neq 0$ {\em both} the metallic and the insulating solution are
locally stable, but the insulating one becomes unstable as
$T\rightarrow 0$, in agreement with our stability considerations. 
We also note that as the temperature is increased, the free energy 
of the metallic solution ($\ell =1$) increases, until the {\em
spinodal} is reached, where the local minimum becomes an
inflection point and becomes even locally unstable. Of course, 
this instability is {\em preempted } by a first order transition,
which in this case happens at finite temperature, in agreement with
findings of Refs. \cite{Marcelo3,GeoKra}.

\acknowledgements We thank S. Barle, G. Kotliar, M. Rozenberg, and
Q. Si for useful discussions.  The work at Rutgers was supported by
the NSF under grant DMR 92-24000 and the ONR under grant
N-11378-RUCKENSTEIN. V. D.  was supported by the Alfred P. Sloan
Foundation, by the NSF under Cooperative Agreement No. DMR95-27035,
and the State of Florida.


\begin{references}
\bibitem{gk2} In recent work, Kotliar and Georges (private
communication) have proposed a cluster $1/d$ expansion 
of the $d=\infty$ mean-field theory, that can be interpreted in 
terms of one and two-impurity models. Although this theory cannot
be considered to be a exactly solvable 
limit of a definite lattice model, it offers an elegant way of
restoring the translational invariance in the cluster formulation. 
On physical as well as on technical grounds, we expect the results 
of such a formulation to be close to the findings presented in this
paper.
\end{references}
\end{document}